\newcommand{\A}{{\frak A}} 
\newcommand{\V}{{\cal V}}
\newcommand{\C}{{\Bbb C}}

\newcommand{\R}{{\Bbb R}}
\newcommand{\Cl}{{\frak Cl}}
\newcommand{\Id}{\openone}
\newcommand{\e}{{\frak e}}
\newcommand{\es}{\e^\sigma}
\newcommand{\xv}{\vec{x}}
\newcommand{\xc}{{\bf x}}
\newcommand{\yc}{{\bf y}}
\newcommand{\Spin}{{\rm Spin}}
\newcommand{\Spoin}{{\rm Spoin}}
\newcommand{\G}{{\mit\Gamma}}
\newcommand{\un}{{\frak u}}
\newcommand{\Dir}{{\frak D}}
\def\hc{\dag}
\newcommand{\n}[1]{{\cal N}{(#1)}}

\documentstyle[aps,pra,multicol,amssymb,eqsecnum]{revtex}
\begin{document}
\draft

\title{Clifford algebras and universal sets of quantum gates}
\author{Alexander Yu. Vlasov\cite{byline}}
\address{Federal Radiological Center (IRH),
197101, Mira Street 8, St.-Petersburg, Russia}
\date{19 October 2000}
\maketitle
\begin{abstract}
 In this paper is shown an application of Clifford algebras to the
construction of computationally universal sets of quantum gates for $n$-qubit
systems. It is based on the well-known application of Lie algebras together
with the especially simple commutation law for Clifford algebras, which
states that all basic elements either commute or anticommute.
\end{abstract}
\pacs{PACS numbers: 03.67.Lx, 03.65.Fd, 02.20.Sv}

\begin{multicols}{2}

\section{Introduction}

In this paper is discussed an algebraic approach to the construction of
computationally universal sets of quantum gates. A quantum gate $U$ for a
system of $n$ qubits is a unitary
$2^n \times 2^n$ matrix. It is possible to write $U = e^{i H}$, where
$H$ is the Hermitian $2^n \times 2^n$ matrix.

A set of quantum gates $U_k$ is {\em (computationally) universal} if any
unitary matrix
can be obtained with given precision as a product of matrices $U_k$.
Algebraic conditions of universality can be described using the Lie algebra
$\un$ of the Lie group of unitary matrices \cite{DPDV2,DeuUn}: if there is a
set of Hermitian matrices $H_k$ and if it is possible to generate a basis of
space of all Hermitian $2^n \times 2^n$ matrices using only the
commutators $i[H,G] \equiv i(HG-GH)$, then $U_k = \exp({i \tau H_k})$ are
a universal set of quantum gates if $\tau$ is small enough.

In this paper is presented an alternative approach to the construction of
a universal set of gates using both Lie and Clifford algebras. It is possible
because the algebra $\C(2^n \times 2^n)$ of all $2^n \times 2^n$ complex
matrices is the complex Clifford algebra with $2n$ generators, i.e., there
are $2n$ matrices $\G_k$ with the property
$\{\G_k,\G_l\} \equiv \G_k \G_l + \G_l \G_k = 2\delta_{kl}\Id$
(where $\Id$ is unit matrix) and $2^{2n}$ different {\em products} of $\G_k$
generate a basis of $\C(2^n \times 2^n)$ \cite{VlaTmr,ClDir}.

The $2n$ matrices $\G_k$ are not enough for
proof of universality, because we may not use arbitrary {\em products}
of $\G_k$, but only {\em commutators}. In this paper it is shown
that by using commutators of $\G_k$, it is possible to generate only the
$(2n^2+n)$-dimensional subspace, but it is enough to add only one element
$\G_u$ and the new set is universal, i.e., it generates a full
$4^n$-dimensional space $\un(2^n)$.

All $2n$ matrices $\G_k$ may be chosen to be Hermitian and the full complex
algebra was used for simplification.
The extra Hermitian matrix is $\G_u=i\G_{123} \equiv i \G_1 \G_2 \G_3$
or $\G_{1234}$, or any such product of three or four
different $\G_k$.

A constructive proof of universality using the language of the Clifford
algebras is based on a simple commutation law of $4^n$ basic elements:
they either commute or anticommute, because any such element is a product
of up to $2n$ $\G_k$. Direct construction of any $2^n\times 2^n$ matrix
$\G_I\equiv\prod_{k \in I}\G_{k}$ of the Clifford basis by commutators
of $2n+1$ initial elements is shown below in Sec.~\ref{LACA}, theorem 1.

The question about universality is widely investigated \cite{DPDV2,DeuUn,%
DeuTur,DeuGate,BCDMSW,CleveTMR,UnSim}, but the method discussed in the
present work has some special properties. Construction of a universal set
of gates uses {\em only} infinitesimal and continuous symmetries of group
U$(2^n)$ and does not require such discrete operations as permutations of
qubits or basic vectors related to the ``classical limit of quantum circuits.''
The properties of discrete, binary transformations of qubits simply emerge
here from the structure of infinitesimal transformations of Hilbert space,
i.e., directly from Hamiltonians, cf. \cite{DPDV2,UnSim}.

\section{Clifford algebras}

\subsection{General definitions}\label{gendef}

For $n$-dimensional vector space with a quadratic form (metric)
$g(\xv)$, the Clifford algebra $\A$ is a formal way to represent a square root
of $-g(\xv)$ \cite{VlaTmr,ClDir} or, more formally, $-g(\xv)\Id$ where
$\Id$ is the unit of algebra $\A$. The vector space corresponds to
the $n$-dimensional subspace $\V$ of $\A$:
$\xv \mapsto \xc \equiv \sum_{l=0}^{n-1}{x_l \e_l}$, where
$\xc$, $\e_l \in \V \subset \A$. From $\xc^2 = -g(\xv)$, i.e.,
$ \left(\sum_{l=0}^{n-1}{x_l \e_l}\right)^2 =
 \sum_{i,j=0}^{n-1}{g_{ij} x_i x_j}$
follow the main properties of the generators $\e_l$ of the Clifford
algebra:
\begin{equation}
 \{\e_i,\e_j\} \equiv \e_i \e_j + \e_j \e_i = -2g_{ij}.
\label{ClDef}
\end{equation}

Let $g_{ij}$ be diagonal and $g_{ii} = \pm 1$ (the case $g_{ii} = 0$
is not considered here, but see \cite{VlaTmr}).
Then,
\begin{mathletters}
\label{ClMink}
\begin{eqnarray}
 \e_i \e_j &=& -\e_j \e_i \quad (i \neq j),
\label{ClMink:1} \\
 \e_i^2 &=& \pm 1.
\label{ClMink:2}
\end{eqnarray}
\end{mathletters}

It is clearer from Eq.~(\ref{ClMink}) that it is possible to generate no more
than $2^n$ different products of up to $n$ $\e_i$. A linear span of all the
products is a full algebra $\A$ \cite{ClDir}. Let us use the notations
$\e_I = \e_{i_1 i_2 \ldots i_k} \equiv \e_{i_1}\e_{i_2}\cdots\e_{i_k}$, where
$k$ is the number of multipliers or {\em the order} of $\e_I$, $k = \n I$.

If there are no algebraic relations other than Eq.~(\ref{ClMink}), then
the algebra has a maximal dimension $2^n$ and is called the {\em universal}
Clifford algebra, $\Cl(g)$, because for any other Clifford algebra $\A$
with the same metric $g(\xv)$ there is a homomorphism $\Cl(g) \to \A$
(see Ref. \cite{ClDir}).

Let us use the notation $\Cl(l,m)$ for the diagonal metric Eq.~(\ref{ClMink})
with $l$ pluses and $m$ minuses in Eq.~(\ref{ClMink:2}), i.e., for
pseudo-Euclidean (Minkowski) space $\R^{l,m}$. There is a special
notation for Euclidean space: $\Cl(n) \equiv \Cl(n,0)$ and
$\Cl_+(n) \equiv \Cl(0,n)$.

Complexification of any Clifford algebra $\Cl(l,m)$ with $l+m=n$ is
the same complex algebra $\Cl(n,\C)$, because all signs in
Eq.~(\ref{ClMink:2}) may be ``adjusted'' by the substitution $\e_k \to i\e_k$.

Let us denote $\es_I \equiv \sqrt{\e_I^2}\e_I$, i.e.,
if $\e_I^2 = 1$, then $\es_I=\e_I$, but if $\e_I^2 = -1$,
then $\es_I=i\e_I$ and so always $(\es_I)^2 = 1$.

\subsection{Matrix representations}

All complex Clifford algebras in even dimension $\Cl(2n,\C)$ are
isomorphic with a full algebra of $2^n \times 2^n$ complex matrices
\cite{VlaTmr,ClDir}. The simplest case $\Cl(2,\C)$ is the Pauli algebra.
Matrices $\sigma_x$ and $\sigma_y$ can be chosen as generators
$\e_0$, $\e_1$ and $\sigma_z$ is $i\e_0\e_1 = \es_{01}$.

The Pauli algebra is four-dimensional complex algebra and
can also be considered as eight-dimensional real algebra,
$\Cl_+(3)$. Prevalent applications of Clifford algebras in the theory
of NMR quantum computation \cite{Cory,Havel} are based on a real
representation $\Cl_+(3)$ rather than on a complex one $\Cl(2,\C)$, discussed
in the present work. These two approaches are very close, but may be
different in some of the details.

There is simple recursive construction of the complex Clifford algebra
with an even number of generators $\Cl(2n,\C)$ with $\Cl(2,\C)$.
For $n=1$, it is the Pauli algebra, and if there is some algebra
$\Cl(2n,\C)$ for $n \geq 1$, then
\begin{equation}
\Cl(2n+2,\C) \cong \Cl(2n,\C) \otimes \Cl(2,\C).
\label{rec2n}
\end{equation}
The proof of Eq.~(\ref{rec2n}) is as follows: if $\e_0,\ldots,\e_{2n-1}$ are
$2n$ generators of $\Cl(2n,\C)$ then $\Id_{2n} \otimes \e_0$ and
$\Id_{2n} \otimes \e_1$ together with $2n$ elements $\e_k \otimes \es_{01}$
are $2n+2$ generators of $\Cl(2n+2,\C)$.

Direct construction of $\Cl(2n,\C)$ is \cite{VlaTmr,ClDir}
\begin{mathletters}
\label{genClnnC}
\begin{eqnarray}
 \G_{2k} & = &
 {\underbrace{\Id\otimes\cdots\otimes\Id}_{n-k-1}}\otimes
 \sigma_x\otimes\underbrace{\sigma_z\otimes\cdots\otimes\sigma_z}_k \, , \\
 \G_{2k+1} & = &
 {\underbrace{\Id\otimes\cdots\otimes\Id}_{n-k-1}}\otimes
 \sigma_y\otimes\underbrace{\sigma_z\otimes\cdots\otimes\sigma_z}_k \, ,
\end{eqnarray}
\end{mathletters}
with $\e_l \corresponds \G_l$, $\e_l^2 = \Id$, $\forall\,l \in 0,\ldots,2n-1$.
More generally, algebraic properties of elements $\e_l$ used in the paper are
the same for different matrix representations $\e_l \corresponds M \G_l M^{-1}$,
where $M \in {\rm SU}(2^n)$.

\subsection{Spin groups}\label{spin}

Most known physical applications of Clifford algebras are due to
{\em spin groups}. The group has 2:1 homomorphism with an orthogonal (or
pseudo-orthogonal) group and is related to the Dirac equation \cite{ClDir} and
the transformation properties of wave functions in quantum mechanics.

Each element $\xc \in \V$ (see the definition of $\V$ above in
Sec.~\ref{gendef}) has an inverse $\xc^{-1} = -\xc/g(\xv)$ if
$g(\xv) \ne 0$. All possible products of {\em even} number of such elements
with $|g| = 1$ is the {\em spin group}. It is $\Spin(n)$ for $\Cl(n)$
and for $\Cl_+(n)$. The group has
2:1 homomorphism with SO$(n)$. For $s \in \Spin(n)$ an element of SO$(n)$
is represented as $r_s \colon \xc \mapsto s \xc s^{-1}$ \cite{ClDir}.

Because only products of an even number of elements of $\Cl(n)$ are used
in the definition of $\Spin(n)$, the group is a subset of even subalgebra
$\Cl^e(n) \subset \Cl(n)$. In the Euclidean case, $\Cl^e(n)$ is isomorphic
with $\Cl(n-1)$ and due to the property $\Spin(n+1)$ may be defined as a
subset of $\Cl(n)$.

Construction of the $\Spin(n+1)$ group from $\Cl(n)$ is
sometimes called the {\em spoin group} \cite{ClDir},
$\Spoin(n) \cong \Spin(n+1)$.

Let us consider $(n+1)$-dimensional space $\lambda \Id \oplus \V$,
i.e., combinations $\yc = \lambda + \xc$, $\xc \in \V$.
Let $\Delta(\yc) \equiv \lambda^2 + g(\xv)$. The elements have
an inverse $(\lambda + \xc)^{-1} = (\lambda - \xc)/\Delta(\yc)$ if
$\Delta(\yc) \ne 0$. Products of {\em any} number of such elements with
$|\Delta| = 1$ is the $\Spoin(n)$ group \cite{ClDir}.

The group $\Spoin(n)$ is 2:1 homomorphic with SO$(n+1)$. For
$s \in \Spoin(n)$, an element of SO$(n+1)$ is represented as
$r_s \colon \yc \mapsto s \yc (s')^{-1}$, where
$\yc = y_n + \sum_{l=0}^{n-1} y_l \e_l$ and ($'$) is the algebra automorphism
defined with basis elements as $\e_I'=(-1)^{\n I}\e_I$ \cite{ClDir}.

\subsection{Lie algebras and Clifford algebras}\label{LACA}

Clifford algebra is Lie algebra with respect to a bracket operation
$[a,b] \equiv a b - b a$ \cite{ClDir}.
Here we prove a result that is necessary for the construction of a
universal set of gates.

{\em Theorem 1.} Let $\Cl(n,\C)$ be the Clifford algebra and $n$ be even.
There are enough $n$ generators $\e_k$, $k=0,\ldots,n-1$ and any
element $\e_I$ with $\n I=3$ or $\n I=4$ to generate elements of any order
only using commutators of these $n+1$ elements.

A proof of this result has several steps.

(i) If there are $n$ elements $\e_0,\ldots,\e_{n-1}$, it is possible by
using commutators to generate also all elements of second order, i.e.,
$[\e_i,\e_j] = 2\e_i\e_j \equiv 2\e_{ij}$.

(ii) If there are all elements of second order and an element of third order,
for example $\e_{012}$, it is
possible to generate any element of third order, i.e.,
$2\e_{01m} = [\e_{012},\e_{2m}]$, $2\e_{0nm} = [\e_{01m},\e_{1n}]$,
$2\e_{pnm} = [\e_{0nm},\e_{0p}]$.

(iii) Analogously, if there is any element of order $2k+1$, it is possible
to generate any element of the same order using no more than $2k+1$
commutators with elements $\e_{ij}$.

(iv) If we have all elements of third order, it is possible to generate any
element of fourth order, $2\e_{ijkl} = [\e_{ijk},\e_l]$.

(v) Analogously, if we have all elements with the order $\n I=2k+1$, it is
possible to generate any element of order $2k+2$,
$2\e_{I \cup l} = [\e_I,\e_l]$, where $l \notin I$.

(vi) If we have an element of fourth order, it is possible to generate some
element of third order, $2\e_{ijk} = [\e_{ijkl},\e_l]$ (and so any
element of third and fourth order).

(vii) Analogously, if we have an element of order $2k+2$, it is possible to
generate some element of order $2k+1$ [and so any element with the order
$2k+1$ or $2k+2$ as in the steps (iii) and (v)].

(viii) We have all elements with order less than or equal to $2k$, $k \ge 2$ due
to steps (i), (ii), and (iv) and we can prove the theorem by recursion: by using
a commutator of an element with order $2k-1$ and an element with order 3 it is
possible to generate an element of order $2k+2$ and so any elements of order
$2k+1$ or $2k+2$, as in the step (vii). 

{\em Note 1.} Instead of elements $\e_0,\ldots,\e_{n-1}$,
it is possible to use $\e_0$ together with $n-1$ elements $\e_{l-1,l}$:
$[\e_0,\e_{01}]~=~2\e_1, \ldots, [\e_{l-1},\e_{l-1,l}] = 2\e_l$.

{\em Note 2.} If $n$ is odd, it is impossible to generate only an element with
the order $n$, because due to step (vii) of recursion it would be generated only
from an even element with the order $n+1$, but there are no such elements. So
in this case we need $n+2$ elements, the extra one being $\e_{0,\ldots,n-1}$.

{\em Note 3.}
If we use only $n$ generators $\e_i$, then
together with $n(n-1)/2$ commutators $[\e_k,\e_j] = 2\e_{kj}$, $k \ne j$,
it is possible to generate $n+n(n-1)/2 = n(n+1)/2$ elements, because, as
may be checked directly, any new commutators may not generate an element
with order more than 2. It is the Lie algebra of the $\Spoin(n)$ group,
because products of $\exp(\epsilon \e_k) \approx \Id + \epsilon \e_k$
belong to that group and the dimension of the group is the
same, $\dim\Spoin(n) = \dim {\rm SO}(n+1) = n(n+1)/2$. Despite the fact that
only elements $\e_I$, $\n I \le 2$ belong to the {\em Lie algebra},
all $4^n$ elements $\e_I$, $\n I \le n$ of $\Cl(n)$ belong to the
{\em Lie group} $\Spoin(n)$ by definition and so a linear span of these
elements is the full Clifford algebra.

{\em Note 4.}
The theorem was proved rather for the more general case of the Lie algebra of
the complex Lie group GL$(N,\C)$, $N=2^{n/2}$ of all matrices $M$,
$\det(M)\neq 0$, than for the unitary group
${\rm U}(N) \subset {\rm GL}(N,\C)$. The proof for the Lie algebra $\un(N)$
of the unitary group U$(N)$ is directly implied. It is sufficient to choose
the initial matrices in $\un(N)$ for a given representation, after which the
Lie brackets may produce only matrices in $\un(N)$ for each step of the proof.

It should be mentioned that there are two traditions for representations
of $\un(N)$. In physical applications, Hermitian matrices $H$ are used,
the Lie brackets are $i[a,b]$, and the unitary matrices are represented
as $U = \exp(-i\tau H)$ due to relations with Hamiltonians and a quantum
version of Poisson brackets \cite{Dirac}. In Eq.~(\ref{genClnnC}),
elements $\e_l = \G_l$, $i\e_{012}$, and $\e_{0123}$
(and $i\e_{kl}$, see {\em Note 1}), i.e., all $\es_I$, are Hermitian.
In more general mathematical applications,
$\un(N)$ are skew-Hermitian matrices $A^\hc=-A$ and ``$i$'' multipliers
are not present in the expressions for the commutators and the exponents
\cite{ClDir}, because $A \corresponds i H$.

\section{Application to quantum gates}

\subsection{Universal set of quantum gates}

Now let us discuss the construction of universal gates more directly. Instead
of Lie algebra $\un(2^n)$, we should work with Lie group U$(2^n)$. Then
an element $\es_I$ corresponds to a unitary gate
$U^\tau_I \equiv \exp(i\tau\es_I)$. One of the advantages of elements $\es_I$
is the analytical expression for the exponent:
\begin{equation}
 U^\tau_I = e^{i\tau\es_I} = \cos(\tau)+i\sin(\tau)\es_I.
\label{sincos}
\end{equation}
Equation~(\ref{sincos}) is valid for any operator with the property $\e^2=\Id$
and it is true for all $4^n$ basis elements $\es_I$.

It is also possible due to Eq.~(\ref{sincos}) to combine the approach with
{\em infinitesimal} parameters $\tau$ \cite{DPDV2,DeuUn} and an approach with
{\em irrational} parameters \cite{DeuTur,DeuGate}. The smaller $\tau$ is, the
higher is the precision in generation of arbitrary unitary gates in
\cite{DPDV2,DeuUn}.
Due to Eq.~(\ref{sincos}), accuracy may be arbitrarily high if
we use gates $U_I = e^{i\varpi\es_I}$ with irrational $\varpi/\pi$ because
for any $\tau$ there exists the natural number $N$ and $\varepsilon < \tau$:
$U^\varepsilon_I = (U_I)^N$. It should be mentioned that the unitary gates do
not necessarily have irrational coefficients even if $\varpi/\pi$ is
irrational, for example $U_I = 0.8 + 0.6 \es_I$.

Yet another advantage of the elements $\es_I$ is a simpler expression for
``commutator gate''. In the usual case \cite{DPDV2,DeuUn}, it is generated as
$$
e^{i\tau i[H_k,H_l]} \approx e^{i\sqrt{\tau}H_k} e^{i\sqrt{\tau}H_l}
e^{-i\sqrt{\tau}H_k} e^{-i\sqrt{\tau}H_l},
$$  
and the expression has precision $O(\tau^{1.5})$.
For elements $\es_I$, there is an exact construction.
If $H_I = \es_I$ and $H_J = \es_J$, then either $[H_I,H_J] = 0$ or
$[H_I,H_J] = 2 H_I H_J$. The first case is trivial and for the second case
due to Eq.~(\ref{sincos}),
$$
e^{i\tau i[H_I,H_J]/2} = e^{-\tau H_I H_J} =
e^{i\case\pi/2 H_I} e^{i\tau H_J} e^{-i\case\pi/2 H_I}.
$$

After construction of the basis of Hermitian matrices $H_I~=~\es_I$, it is
possible to use an expression
\begin{eqnarray*}
e^{\sum_I\alpha_I H_I} &=& \left(e^{\frac1N\sum_I\alpha_I H_I}\right)^N \\
&\approx&\left(\prod e^{\frac1N \alpha_I H_I}\right)^N \equiv
\left(\prod U_I^{\frac{\alpha_I}{N}}\right)^N.
\end{eqnarray*}
The expression has accuracy $O\left(\sum\alpha_I^2/N\right)$.

The approach to a universal set of gates $U$ is more convenient and
constructive if we know the Hermitian matrix $H$, $U^\tau = e^{i\tau H}$.
It is not a principal limitation, because for physical realizations
we should know the Hamiltonian to construct the gates. It is also related
to the universal quantum simulation \cite{UnSim} in which $H$ is the
Hamiltonian and $\tau$ is a real continuous parameter, the time of
``application.''

The description with an exponent may be even more complete, because by using
$H$ it is possible to find a unique $U = \exp(iH)$, but using $U$ it is not
always possible to restore $H$ because there are many $H$'s for the same $U$.
A simple example is $U = i \sigma_\alpha \otimes \sigma_\beta$ with two
arbitrary Pauli matrices:
$
 U = e^{i\pi\,(\sigma_\alpha \otimes 1 + 1 \otimes \sigma_\beta)} =
     e^{i\pi\, \sigma_\alpha \otimes \sigma_\beta}.
$

\subsection{Two-qubit quantum gates}

Let us show how to build a universal set of one- and two-qubit gates
using Eq.~(\ref{genClnnC}). For example, it may be $2n+1$
gates $\exp(i\tau\e_I)$, where $\e_I$ are $\e_0$, $i\e_{l-1,l}$
with $l = 1,\ldots,2n-1$, and $i\e_{012}$:
\begin{mathletters}
\label{twoQG}
\begin{eqnarray}
 \e_{0} & = & \Id^{\otimes (n-1)}\otimes\sigma_x ,
 \label{twoQG:a}\\
 \case 1/i \e_{2k,2k+1} & = & \Id^{\otimes (n-k-1)}\otimes
 \sigma_z\otimes\Id^{\otimes k} ,
 \label{twoQG:b}\\
 \case 1/i \e_{2k+1,2k+2} & = & \Id^{\otimes (n-k-2)}\otimes\sigma_x\otimes
 \sigma_x\otimes\Id^{\otimes k} ,
 \label{twoQG:c}\\
 \case 1/i \e_{012} & = & \Id^{\otimes (n-2)}\otimes\sigma_x\otimes\Id
 \label{twoQG:d}
\end{eqnarray}
\end{mathletters}
with $k = 0,\ldots,n-1$ or $n-2$.
The elements were discussed in {\em Note 1}, and it was shown that they
generate the full Lie algebra $\un(2^n)$.

\subsection{Nonuniversal set of quantum gates}

In \cite{DeuUn}, an interesting question was raised, asking which sets of
gates are {\em not} universal (and why).

Products of gates $U^\tau_k = e^{i \tau \e_k} = \cos(\tau) + i\e_k\sin(\tau)$
generate a group $\Spin(2n+1)\cong\Spoin(2n) \subset U(2^n)$ due to
{\em Note 3}. It is an interesting example of nonuniversality when only one
extra gate like $e^{i \tau \es_{012}}$ may produce a universal set with
``an exponential improvement'' from a subgroup $\dim\Spoin(2n)=n(2n+1)$ to
a full group $\dim U(2^n) = 2^{2n}$.

This result is more important if the extra gate $e^{i \tau \es_I}$
with $\n I = 3$ or $\n I = 4$ has a different physical nature from the
gates with $\n I = 1$ and $\n I = 2$. It is not clear from
Eq.~(\ref{twoQG}) with the extra gate generated by Eq.~(\ref{twoQG:d})
is the simple one-qubit gate. But this is not so for physical systems with
{\em natural} Clifford and spin structure.

A possible reason is the Schr\"odinger equation for $n$ particles without
interaction \cite{dau:qm}:
$
i\hbar({\partial\psi}/{\partial t}) =
\case 1/2 \hbar^2 \sum_{a=1}^n ({\Delta_a}/{m_a}) \psi,
$ 
or using $m_a=m$ and the Laplacian $\Delta_N$ with $N = \nu n$ variables,
it is possible
to write for stationary solutions with total energy $E$,
\begin{equation}
 (\Delta_N + \lambda^2) \psi(x_0,\ldots,x_{N-1}) = 0,
\label{LapLam}
\end{equation}
where $\lambda \equiv \sqrt{2mE}/\hbar$. Let the dimension of one
particle motion be $\nu = 2$ for simplicity, $N=2n$.

Let us consider a full basis $\phi_{\bf p}({\bf x}) \equiv e^{i({\bf p, x})}$
on Hilbert space ${\cal L}$ of wave functions $\psi \in {\cal L}$. Here
${\bf p},{\bf x} \in \R^N$ and $({\bf p, x})$ is the scalar product.
The plane waves $\phi_{\bf p}$ correspond to $n$
particles with definite momenta. If $O \in {\rm SO}(N)$, then a transformation
defined on the basis as $\Sigma_O \colon \phi_{\bf p} \to \phi_{O{\bf p}}$
is a {\em symmetry} of Eq.~(\ref{LapLam}). It is an analog of the classical
transition between two configurations {\em with the same total kinetic energy}
in ``billiard balls'' conservative logic \cite{toff}.

The general Dirac operator \cite{ClDir} is the first-order differential
operator
$\Dir_N = \sum_{i=0}^{N-1}i\e_k({\partial}/{\partial x_k})$ with a property
$\Dir_N^2 = -\Delta_N$. If to use the Dirac operator for factorization
of Eq.~(\ref{LapLam}),
\begin{equation}
 (\Dir_N - \lambda)(\Dir_N + \lambda) {\bf \Psi}(x_0,\ldots,x_{N-1}) = 0,
\end{equation}
then each component of ${\bf \Psi}$ is a solution of Eq.~(\ref{LapLam}) and
the action of the $\Spin(N)$ group on ${\bf \Psi}$ corresponds \cite{ClDir}
to SO$(N)$ symmetry $\Sigma_O$ described above and it has some
analog in the classical physics of billiard balls. A $\Spoin(N)$ group
is represented less directly, but it can be considered as a symmetry between
two stationary solutions with {\em different} total energies.

The example above shows that it is possible to find some classical
correspondence for elements $\e_I$, $\n I = 2$ of the spin group and maybe
for generators $\n I= 1$ of the spoin group, but the special element with
$\n I = 3$ does not have some allusion with classical physics.

\end{multicols}
\end{document}